 \documentclass[apl,aip,preprint,]{revtex4}
 \newcommand{\beq}{\begin{equation}}
\newcommand{\eeq}{\end{equation}}
\usepackage{graphics}
\usepackage{epsf}
\usepackage{color}
\usepackage{amsmath}
\usepackage{amsfonts}
\usepackage{amssymb}
\usepackage{graphicx}
\usepackage{epsfig}
\begin{document}

\title{Metallic Conductance at the Interface of Tri-color Titanate Superlattices}
\author{M. Kareev}
\email{mkareev@uark.edu}
\affiliation  {Department of Physics, University of Arkansas, Fayetteville, Arkansas 72701, USA}
\author{Yanwei Cao}
\affiliation  {Department of Physics, University of Arkansas, Fayetteville, Arkansas 72701, USA}
\author{Xiaoran Liu}
\affiliation  {Department of Physics, University of Arkansas, Fayetteville, Arkansas 72701, USA}
\author {S. Middey}
\affiliation  {Department of Physics, University of Arkansas, Fayetteville, Arkansas 72701, USA}
\author{D. Meyers}
\affiliation  {Department of Physics, University of Arkansas, Fayetteville, Arkansas 72701, USA}
\author{ J. Chakhalian}
\affiliation  {Department of Physics, University of Arkansas, Fayetteville, Arkansas 72701, USA}

\date{\today}

\begin{abstract}

Ultra-thin tri-color (tri-layer) titanate superlattices ([3u.c. LaTiO$_{3}$/2u.c. SrTiO$_{3}$/3u.c. YTiO$_{3}$], u.c.~=~unit cells) were grown in a layer-by-layer way on single  crystal  TbScO$_3$ (110) substrates  by pulsed laser deposition. High sample quality and electronic structure were characterized by  the combination of \textit{in-situ}  photoelectron   and \textit{ex-situ} structure and surface morphology  probes. Temperature-dependent sheet resistance  indicates the presence of  metallic  interfaces in both [3u.c. LaTiO$_{3}$/2u.c. SrTiO$_{3}$] bi-layers and all  the tri-color structures, whereas  a [3u.c. YTiO$_{3}$/2u.c. SrTiO$_{3}$] bi-layer shows insulating behavior. Considering that in the bulk YTiO$_3$ is ferromagnetic below 30~K, the tri-color titanate superlattices provide an opportunity to induce tunable spin-polarization into the two-dimensional electron gas (2DEG) with Mott carriers.

\end{abstract}

\maketitle

 Recent studies of emergent phenomena at the interfaces of  transition metal oxides have attracted a tremendous interest due to  the variety of novel phenomena,\cite{Nmat-2012-Hwang,Nmat-2012-JC,Science-2010-Mannhart,ARCMP-2011-Zubko} which are either hindered or not  attainable in bulk materials including two dimensional electron gas (2DEG),\cite{Nmat-2012-Hwang,Science-2010-Mannhart,MRS-2008-Mannhart} spin and orbital  reconstruction at the superconductor/ ferromagnet interface,\cite{NPhys-2006-JC,Science-2007-JC} magneto-electric effects,\cite{NComm-2012-Rogdakis,Nature-2005-Lee} magnetism in layers of non-magnetic materials,\cite{APL-2001-Takahashi} metal-insulator  transitions,\cite{PRL-2012-Liu} anomalous Hall effect\cite{PRB-2011-Kozuka,PRL-2013-Nakayama} to  mention a few. In particularly, the appearance of a  2DEG was observed at the junction between two insulators composed of non-polar band insulator SrTiO$_3$ and polar band insulator LaAlO$_3$.\cite{Nmat-2012-Hwang, Nature-2004-Ohtomo,Nmater-2006-Nakagawa,PRL-2007-Willmott,Science-2006-Thiel,Science-2007-Reyren} Similar metallic interfaces have been reported  in other oxide materials such as LaTiO$_3$/SrTiO$_3$ (LTO/STO),
\cite{Nature-2002-Ohtomo,JJAP-2004-Shibuya,Science-2011-Jang,PRB-2010-Kim,APL-2010-Ohtsuka,PRB-2012-Rastogi,Ncomm-2010-Biscaras,Nmater-2013-Biscaras} LaVO$_3$/SrTiO$_3$,\cite{PRL-2007-Hotta}  LaGaO$_3$/SrTiO$_3$,\cite{APL-2010-Perna}  GdTiO$_3$/SrTiO$_3$,\cite{APL-2011-Moetakef} and Al$_2$O$_3$/SrTiO$_3$.\cite{Ncomm-2013-Chen} On the other  hand, since in complex oxides spin, charge and orbital degrees of  freedom are strongly coupled, it would be  interesting to investigate if the 2DEG can be  spin-polarized (SP),\cite{APL-2011-Moetakef,PRL-2012-Reyren} possibly  opening  new opportunities for spintronics applications.\cite{RMP-2004-Zutic} Recent, theoretical studies have predicted  the possibility of SP-2DEG at the interfaces of SrMnO$_3$/LaMnO$_3$ and LaAlO$_3$/EuO.\cite{PRL-2008-Nanda,PRB-2009-Wang} Based on this,  designing artificial hetero-junctions of  correlated  oxides synthesised with unit-cell precision\cite{Nature-2002-Ohtomo,Science-2011-Jang,Nmater-2006-Nakagawa,PRL-2007-Willmott} is a key step towards, these emerging physics and new nano-devices.

In this letter, we report growth  and transport properties of tri-color titanate superlattices [3u.c. LaTiO$_{3}$/2u.c. SrTiO$_{3}$/3u.c. YTiO$_{3}$]$_4$ (simplified as [3LTO/2STO/3YTO]) and bi-color titanate superlattices [3u.c. LTO/2u.c. STO]$_4$ ([3LTO/2STO]) and [3u.c. YTO/2u.c. STO]$_4$ ([3YTO/2STO]) (see Fig.~1~(a)). For tri-color structures, from the bottom the stacking sequences were ABCBABCBA , and A~=~3YTO, B~=~2STO, C~=~3LTO, respectively. In the bulk, LaTi$^{3+}$O$_3$ is a Mott insulator with 3\textit{d}$^{1}$ electron configuration and G-type antiferromagnetic (AFM) ground state below$<$ 146~K,\cite{PRB-2003-Cwik} whereas the Mott insulator YTi$^{3+}$O$_{3}$ (YTO) with 3\textit{d}$^{1}$ electron configuration undergoes a ferromagnetic transition below 30~K.\cite{APL-2006-Chae} SrTi$^{4+}$O$_3$ with 3\textit{d}$^{0}$ electronic configuration is a band-gap paramagnetic insulator.  The high quality of surface and crystal morphology and electronic structure are confirmed by the combination of reflection-high-energy-electron-diffraction (RHEED), atomic force microscopy (AFM), X-ray diffraction (XRD), and \textit{in-situ} X-ray photoemission (XPS)\ and Auger spectroscopies. In sharp contrast to  the [3YTO/2STO] superlattice,  for both [3LTO/2STO] and [3LTO/2STO/3YTO] films, DC conductance shows metallic character at all  temperatures below 300~K. By adding ferromagnetic YTO layers into the well-studied LTO/STO superlattice,\cite{Nature-2002-Ohtomo,JJAP-2004-Shibuya,Science-2011-Jang,PRB-2010-Kim,APL-2010-Ohtsuka,PRB-2012-Rastogi,Ncomm-2010-Biscaras,Nmater-2013-Biscaras} a new tri-color titanate superlattice LaTiO$_3$/SrTiO$_3$/YTiO$_3$ (LTO/STO/YTO) is formed. In this structure, because of small  thickness of the STO layer (2 and 3 u.c.), conducting carriers of  2DEG formed at the LTO/STO\ interface are able to interact with magnetic moments of  the YTO layer thus opening a route to the spin polarized version of 2DEG.

 All films in this work were grown on (110) surface (in orthorhombic notation) of TbScO$_{3}$ (TSO) single crystal substrates by pulsed laser deposition   using a KrF excimer laser operating at $\lambda$~=~248~nm and 2~Hz pulse rate with $\sim$2~J/cm$^{2}$ fluence. TSO is selected for two reasons, (1) bulk TbScO$_{3}$ is a insulator without Ti in its chemical  formula, and (2)  bulk TSO has an orthorhombic structure with lattice constant closely  matching  the active titanate layers (see Table I).  The growth is monitored by \textit{in-situ} high-pressure RHEED. During growth, the temperature of the substrate is held at $950\,^{\circ}{\rm C}$; we report heater temperature. Since oxygen deficient  LTO\  and STO\ structures can  easily form other  chemical  phases at high oxygen pressure,  e.g. La$_{2}$Ti$_2$O$_7$ and Y$_{2}$Ti$_2$O$_7$,\cite{APL-2006-Chae,APL-2002-Ohtomo} low oxygen pressure (1~-~3 $\times$ 10$^{-6}$ Torr) was  maintained during the deposition in this work. After the growth, all samples were cooled at about $15\,^{\circ}{\rm C}$/min rate, keeping oxygen pressure constant. Fig.~1~(c)-(e) shows a  sequence of RHEED\ images for a tricolor superlattice [3 LTO/2 STO/3 YTO] upon completion of a specific layer in the junction.  The sharp RHEED patterns with  expected orthorhombic symmetry  indicate  the layer-by-layer two dimensional (2D) growth mode with  proper epitaxial relations among the layers and the substrate. In addition, smooth surface topography of the films was confirmed by AFM; the obtained average surface roughness was found to be \textless\ 120~pm. Structural quality, tensile strain and proper epitaxy  of the superlattices are further confirmed by X-ray Diffraction (XRD) using Cu K$_{\alpha}$ radiation shown in Fig.~1~(f). DC transport properties were measured  in van der Pauw geometry by a Physical  Properties Measurement System (PPMS)\ operating in high resolution mode.

To assure that emergent phenomena arises from the interface and not secondary chemical  phases or defects, electronic structure and chemical composition of the films was further investigated by \textit{in-situ} X-ray photoemission and Auger (not shown) spectroscopies; no discernable impurity signal was observed. On the other hand, as shown in Fig.~2, in three reference films 20 u.c. LTO, 20 u.c. STO, and 25 u.c. YTO (the same growth conditions as for  tri-color superlattice and TSO substrate), the two main Ti peaks marked as Ti 2p$_{1/2}$ and Ti 2p$_{3/2}$ for Ti$^{3+}$ and Ti$^{4+}$ are clearly seen. The peak separation between Ti$^{3+}$ and Ti$^{4+}$ is about 2 eV, showing good agreement with previously reported data.\cite{PRB-2013-Drera} The overall line-shape of the three reference films affirms that after growth the LTO, STO, and YTO layers maintain proper  bulk-like valence states of Ti. At the  same time, for [3LTO/2STO/3YTO], the features of both Ti$^{3+}$ and Ti$^{4+}$ are observed simultaneously as deduced from the titanium binding energy positions  in the reference films.

Next we turn our attention to the transport properties of the tri-color  superlattices. As shown in Fig.~1, there are two distinct interfaces separated by STO layers, namely,  LTO/STO and YTO/STO. The  well-studied LTO/STO interface is known to  be metallic and represents a prototypical  2DEG system;\cite{Nature-2002-Ohtomo,JJAP-2004-Shibuya,Science-2011-Jang,PRB-2010-Kim,APL-2010-Ohtsuka,PRB-2012-Rastogi,Ncomm-2010-Biscaras,Nmater-2013-Biscaras} to-date, transport properties of  the  YTO/STO interface have not  been reported.  To understand the conducting  behavior of  the tri- and bi-color  samples we first compare their transport  behavior to  the bulk-like reference samples  of 20 u.c. LTO, 20 u.c. STO and 25 u.c. YTO  grown on TSO. All  those single composition layers demonstrate  characteristic insulating behavior starting  from 300 K (not shown); this  result excludes the effects of defects and/or oxygen doping in LTO, STO, and YTO layers  during the tri-color superlattice growth. We also point  out, that this observation is particularly  important for  LTO layers, since  in the bulk LTO can be  easily  oxygen over-doped to  form metallic LaTiO$_{3+\delta}$ .\cite{PRB-1999-Taguchi} Moreover, in Fig. 1(c) a typical RHEED pattern observed during the LTO growth is shown, yielding well-defined spots and streaks without twofold superstructure. In the case of oxygen overdoping, the appearance of such superstructure peaks is expected, since they would originate from the c-axis doubling in the perovskite unit cell, as previously reported for LaTiO$_{3.5}$.\cite{PRB-2001-Seo}

After the expected conducting  behavior of individual constituent  layers was established, we proceed with  study of the of tricolor superlattices. Figure 3~(a) demonstrates the  temperature  dependence of  [3 LTO/2 STO/3 YTO]. As immediately  seen,  unlike individual  layers the tri-layer structure  displays  \textit{metallic}  behavior at the wide temperature range from RT down to  2~K. To elucidate the origin of metallicity in this structure we investigate conducting  properties of individual YTO/STO and LTO/STO  in the complimentary bi-color samples (Fig.~3). As seen in Fig.~3~(b) sheet resistance of the [3YTO/2STO] interface is high $\sim$ 30~k$\Omega$  at  room temperature (RT) and rapidly increases with lowered temperature to $\sim$ 514~k$\Omega$ at 70~K; this result  confirms the insulating character of the YTO/STO\ interface.  It is interesting  to  note that this behavior is also different from the bulk Y$_{1-x}$Ca$_{x}$TiO$_{3}$ material which is metallic for $0.4<x<0.8$.\cite{PRB-1993-Taguchi,PRB-1996-Morikawa} In contrast to  the doped bulk  material, the insulating character of the YTO/STO interface is consistent with the theory predicted  band gap for a monolayer of YO buried in STO layers.\cite{Science-2011-Jang} Based on this result and the fact that the LTO/STO samples demonstrate distinct metallic behavior, we conclude that conductivity in the  tri-color superlattice mainly originates from the LTO/STO interface. Furthermore,  assuming that the conducting channels are homogenious, the averaged (a bulk-like resistivity $\rho$ (in ohm per cm) can be calculated by multiplying the sheet resistance by the film thickness in cm) resistivity of [3LTO/2STO] reaches $\sim$ 227~$\mu\Omega\cdot$cm very  similar to  the previously  reported value of  200~$\mu\Omega\cdot$cm in the best [2LTO/3STO] samples.\cite{JJAP-2004-Shibuya} In comparison to  this, as shown in Fig.~3~(a), sheet resistance of tricolor films have similar temperature  dependence albeit with slightly larger value of resistance. This increasing of residual resistance is expected  and  likely arises from the enhanced scattering of conducting carriers by the YTO/STO interface.

In summary, we have developed layer-by-layer growth of high quality tri-color and bi-color titanate superlattices. The crystal and electronic structures of the film are confirmed by XRD, AFM, XPS and Auger spectroscopies. Unlike constituent  layers which  are all insulators, the tri-color superlattices demonstrate  distinct metallic behaviour at wide range of temperatures. A comparison of the bi-color  [3YTO/2STO] and  [3LTO/2STO] seems confirm that metallicity stems from the LTO/STO\ interface leaving  YTO/STO\ insulating. This  heterojunctions with the  two  dissimilar Mott/band gap interfaces may provide a  another route to induce tunable spin-polarization (by controlling thickness of STO layer in tri-color structures) into 2DEG of the LTO/STO\ interface to form a convenient platform for novel spintronics applications. By tuning thickness of SrTiO$_3$ layer in tri-color structures, the spin-polarization may be controlled. Retaining the intrinsic magnetic field for spin accumulation, without any gate voltage applied,\cite{Science-2005-Inoue} may enable design of spintronic devices, which could exploit the large Rashba coupling in 2DEG.

\begin{acknowledgments}

JC is supported by DOD-ARO Grant No. 0402-17291.
We deeply acknowledge discussions with Christsos Panagopolous.
\end{acknowledgments}

\newpage

\begin{figure}[htp]
\includegraphics[width=0.6\textwidth]{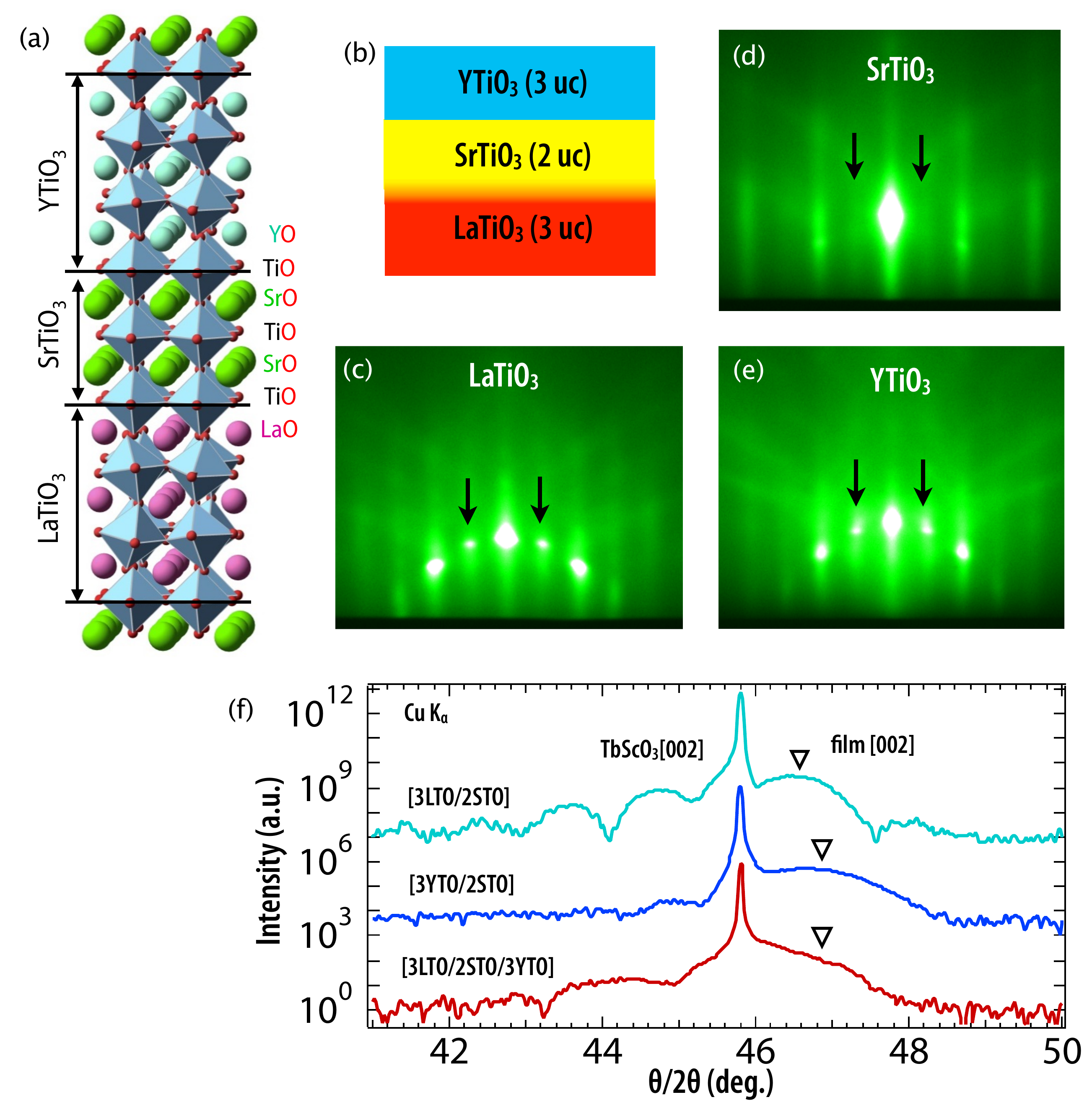}
\caption{\label{} (Color online) (a) and (b) possible crystal structures and
block sketch of growth sequences of tricolor titanate superlattices, respectively.
(c)-(e) RHEED patterns for LaTiO$_3$, SrTiO$_3$, and YTiO$_3$ layers, respectively, during
growth. Black arrows indicate half-order-peaks. (f) XRD data of bi-color film ([3LTO/2STO] and
[3YTO/2STO]) and tri-color film ([3LTO/2STO/3YTO]) near [002] rod (pseudocubic
structure). The main substrate peak (2$\theta\sim 45.8^{o}$)
corresponds  to $\thicksim$~3.958~{\AA}. The black triangles suggest broad peak of film.
Since the  lattice parameter of bulk STO and YTO are smaller than that of substrate (see Table I), the superlattices should be under tensile strain.
For all  three films, the thickness fringes were distinct, additionally testifying for flatness of the layers.}
\end{figure}

\newpage
\begin{figure}[htp]
\includegraphics[width=0.6\textwidth]{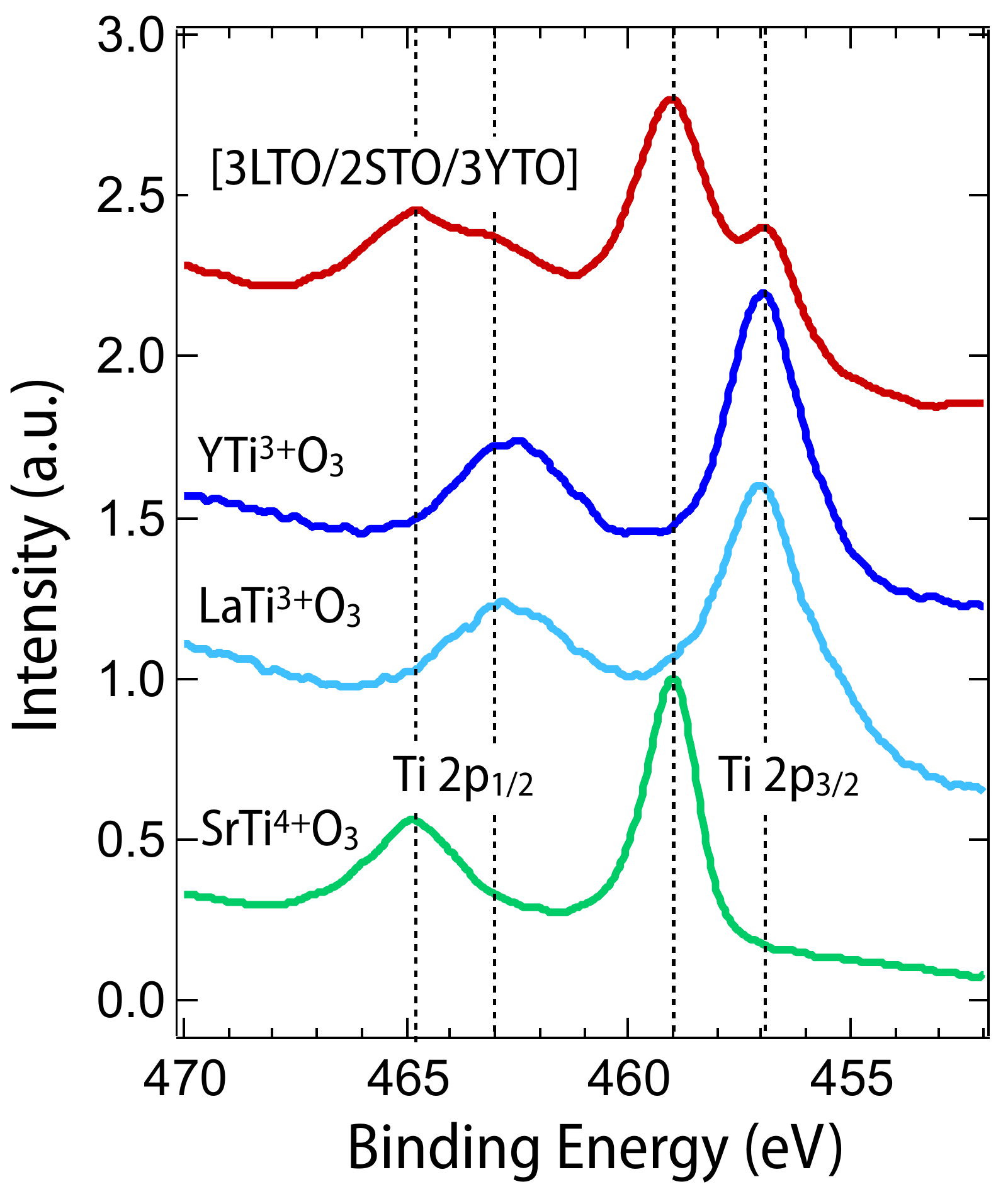}
\caption{\label{} (Color online) XPS spectra (Mg anode) of films with pure
Ti$^{4+}$ (20~uc~STO), pure Ti$^{3+}$ (20~uc~LTO and 25~uc~YTO), and mixed Ti$^{4+}$/Ti$^{3+}$ ([3LTO/2STO/3YTO]), respectively.}
\end{figure}

\newpage
\begin{figure}[htp]
\includegraphics[width=0.6\textwidth]{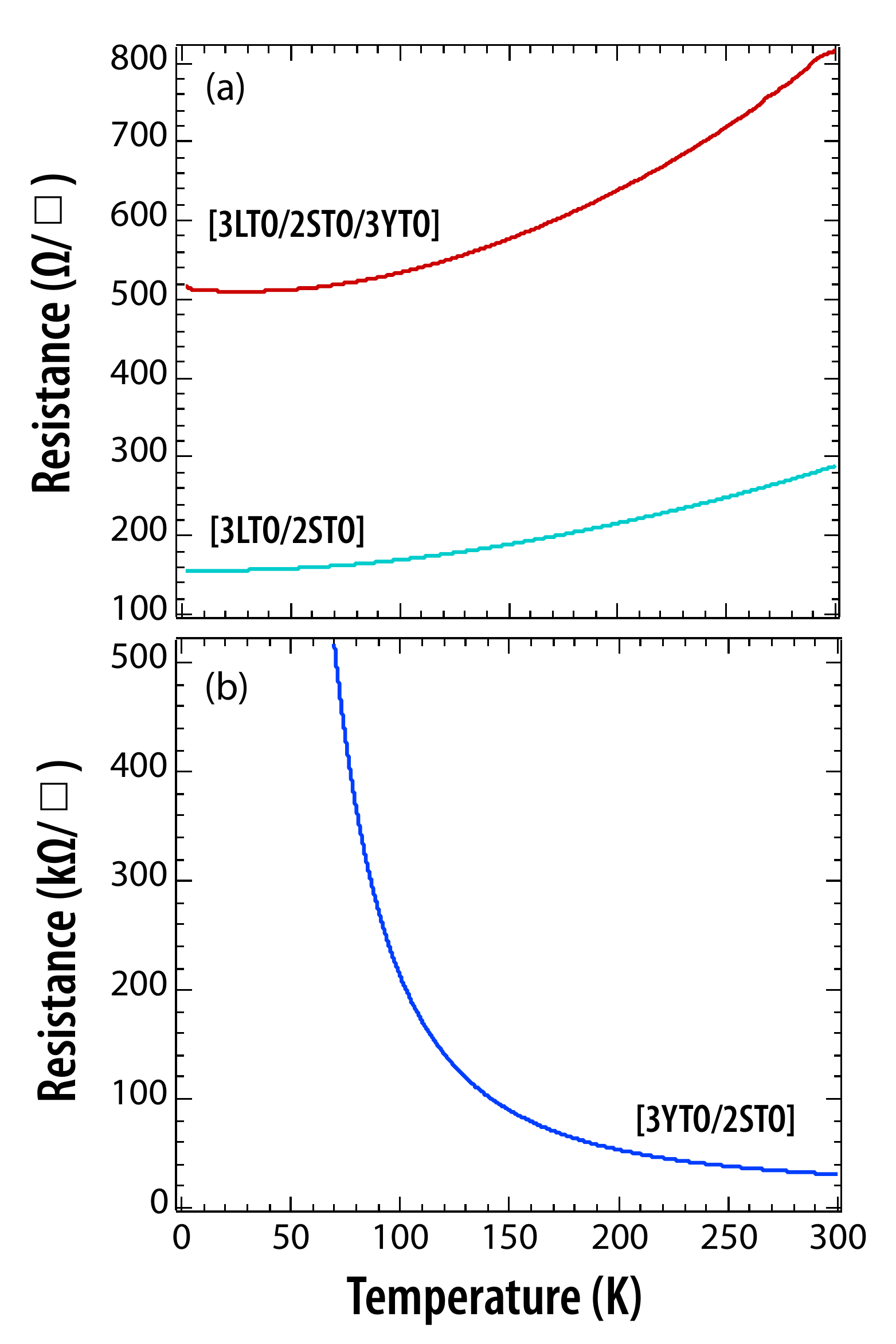}
\caption{\label{} (Color online) Sheet resistance of different titanate superlattices.
(a) Metallic bi-color film [3LTO/2STO]
and tri-color films [3LTO/2STO/3YTO], respectively. (b) Insulated bi-color film [3YTO/2STO].}
\end{figure}

\newpage
\begin{table}
\caption{\label{} Lattice parameters of four perovskite materials used in
this work. a, b, c
are orthorhombic or cubic lattice parameters,\cite{APL-2003-Schubert,Science-2011-Jang}
values of $\sqrt{a^2+b^2}/2$ and c/2 were listed.}
\begin{ruledtabular}
\begin{tabular}{c c c c c c c}
Materials & a({\AA}) & b({\AA}) & c({\AA}) & $\sqrt{a^2+b^2}/2$({\AA}) & c/2({\AA})\\ \hline
\\

TbScO$_3$ & 5.466 & 5.727 & 7.915 & 3.958 & 3.958\\

LaTiO$_3$ & 5.595 & 5.604 & 7.906 & 3.959 & 3.953\\

YTiO$_3$ & 5.341 & 5.686 & 7.621 & 3.901 & 3.811\\

SrTiO$_3$ & 3.905 &  &  & 3.905 &

\end{tabular}
\end{ruledtabular}
\end{table}

\end{document}